\documentclass[%
reprint,
showpacs,
floatfix,
amsmath, 
amssymb, 
aps, 
prl,
]{revtex4-1}
\usepackage{graphicx}
\usepackage{dcolumn}
\usepackage{bm}
\usepackage{natbib}
\usepackage{hyperref}
\usepackage{float}
\usepackage{mathrsfs}
\usepackage{fancyhdr}
\hypersetup{colorlinks=true, citecolor=blue, urlcolor=blue, linkcolor=blue}
\begin{document}
\lhead{}
\chead{}
\rhead{}
\lfoot{}
\cfoot{\thepage}
\rfoot{}

\title{Quantum Photonic Node for On-Chip State Transfer}
\author{Zhaohua Tian}

\author{Pu Zhang}
\email{puzhang0702@hust.edu.cn}

\author{Xue-Wen Chen} 
\email{xuewen\_chen@hust.edu.cn}

\affiliation{School of Physics, Huazhong University of Science and Technology, Luoyu Road 1037, Wuhan, 430074, People's Republic of China}%

\date{\today}
\begin{abstract}
Integrated quantum photonics hold the promise to scale up the system size and form an on-chip quantum network with distributed information processing and simulation units. An outstanding need of such quantum network is to have high fidelity and efficiency on-chip state transfer between distant nodes. Although the nodes are naturally connected via waveguides, it is challenging to fulfill this need because stringent conditions such as spatial mode-matching configuration and time-reversal symmetry have to be satisfied. Here we report a type of quantum photonic nodes consisting of single quantum emitters and cascaded microring resonators for on-chip state transfer. By interfacing the node with a waveguide, we show that all the emission from the node can be funneled into the waveguide and its temporal profile can be synthesized to be time-reversal symmetric. We demonstrate theoretically on-chip quantum state transfer between two distant nodes with near-unity overall success rate can be achieved without any dynamic control. Moreover, we discuss the experimental implementation of our scheme with CMOS compatible integrated photonic platforms and solid-state quantum optics techniques.  
\end{abstract}
\maketitle
While detection, microscopy and spectroscopy of single quantum emitters have become routine in many laboratories, scaling up the system size cooperatively has been of central importance and remained a grand challenge for quantum information science. One viable solution is to develop quantum photonic circuits where a large number of solid-state single quantum systems and optical components can be integrated to one single chip \cite{RN109,RN141,RN107,RN108}. Moreover, compared with free space implementation, photonic circuits in the form of waveguide based optical paths, beam splitters, couplers, interferometers, resonators and so on can provide excellent system stability, far more control over photon's behavior and the interaction with single emitters \cite{RN107,RN112,RN15,RN99,RN105,RN4,RN98,RN20,RN142,RN144}. In quantum photonic circuits, the static quantum nodes, i.e. solid-state quantum emitters, communicate via propagating photons in waveguides and naturally have the potential to form an on-chip quantum network with distributed quantum information processing and simulation units \cite{RN110,RN30}. An outstanding need for this scenario is to have both high fidelity and efficiency state transfer between distant quantum nodes through propagating single photons in waveguide. However, this is an extremely difficult task to accomplish on chip, because it requires all the emission from the sending node should be directed to the receiving node through waveguides in a mode-matched and time-inverted fashion \cite{RN25,RN114,RN37,RN17}. To invert an optical pulse, researchers have recently considered using dynamically modulated cavity arrays \cite{RN24,RN21}, direct modulation with an acousto-optic modulator \cite{RN5}, and heralded single photons generated via four-wave mixing processes of a cold atomic ensemble \cite{RN26,RN27,RN116}. However, despite an impressive progress, inverting a single-photon wave packet without loss for each node on chip remains a formidable challenge, especially as the system size increases.
\par In a wider context of quantum information science, state transfer between distant nodes in free space or discrete configurations has been studied for decades. Various schemes based on dynamic single-photon wave packet shaping, dynamic modulation or adiabatic passage have been proposed for high-fidelity state transfer \cite{RN22,RN23,RN28,RN101,RN115,RN38}. An experiment by Ritter et al. \cite{RN7} implemented the wave packet shaping protocol \cite{RN22} and achieved a fidelity of 84\% and an overall success rate of 0.2\% for the state transfer between two single atoms separately trapped in high-finesse cavities and dynamically controlled by two laser pulses. These protocols generally require specific atomic level schemes and delicate dynamic control during the operation, making their realization on chip at optical frequencies remain elusive.
\par Here we report a conceptually different scheme for on-chip deterministic quantum state transfer between distant nodes through propagating single photons in waveguides. We propose a type of quantum photonic node that consists of a solid-state single quantum emitter, i.e. a two-level system (TLS), and coupled microring resonators (MRRs) for sending or receiving one excitation. The quantum nodes are interfaced by a waveguide and designed in such a way that all the emission of the TLS in the node couples to the waveguide and can be synthesized to have a time symmetric pulse shape. Consequently, we demonstrate theoretically that a deterministic quantum state transfer with near-unity overall success rate can be achieved between two identical quantum photonic nodes without any dynamic control. Moreover, we discuss experimental realizations of our protocol with CMOS compatible integrated photonic technologies.
\par We begin with a general discussion on an ideal quantum node for on-chip state transfer. The aim is to transfer the quantum state or one excitation registered in the emitter of the sending node with ideally unity fidelity and efficiency to the emitter of the receiving node via a single-photon wave packet in waveguide. The goal can be achieved when the emission of the sending node couples with unity-efficiency to the waveguide and simultaneously the emitted single-photon pulse is time-reversal symmetric \cite{RN37,RN17}. The first condition could be satisfied by using well-designed photonic crystal waveguides \cite{RN107}, plasmonic nanowires \cite{RN136}, or plasmonic nanocone structures \cite{RN137}. However, in all the above approaches, the temporal profile of the emission is exponentially decaying because the TLS couples directly and irreversibly to waveguide mode, which is a continuum and has infinite number of degrees of freedom. To circumvent this difficulty, we devise a quantum photonic node, which allows that the TLS couples with a few discrete photon states \cite{RN124} in a cascaded fashion and the channel directly interfacing the waveguide continuum is a result of interference of all involving eigenstates. As explained shortly below, the cascaded coupling scheme provides good control over the amplitude, phase and complex eigenfrequency of each eigenstate of the coupled system, which can be harnessed for single-photon wave packet synthesis. 
\begin{figure}
  \centering
      \includegraphics[width=8.6cm]{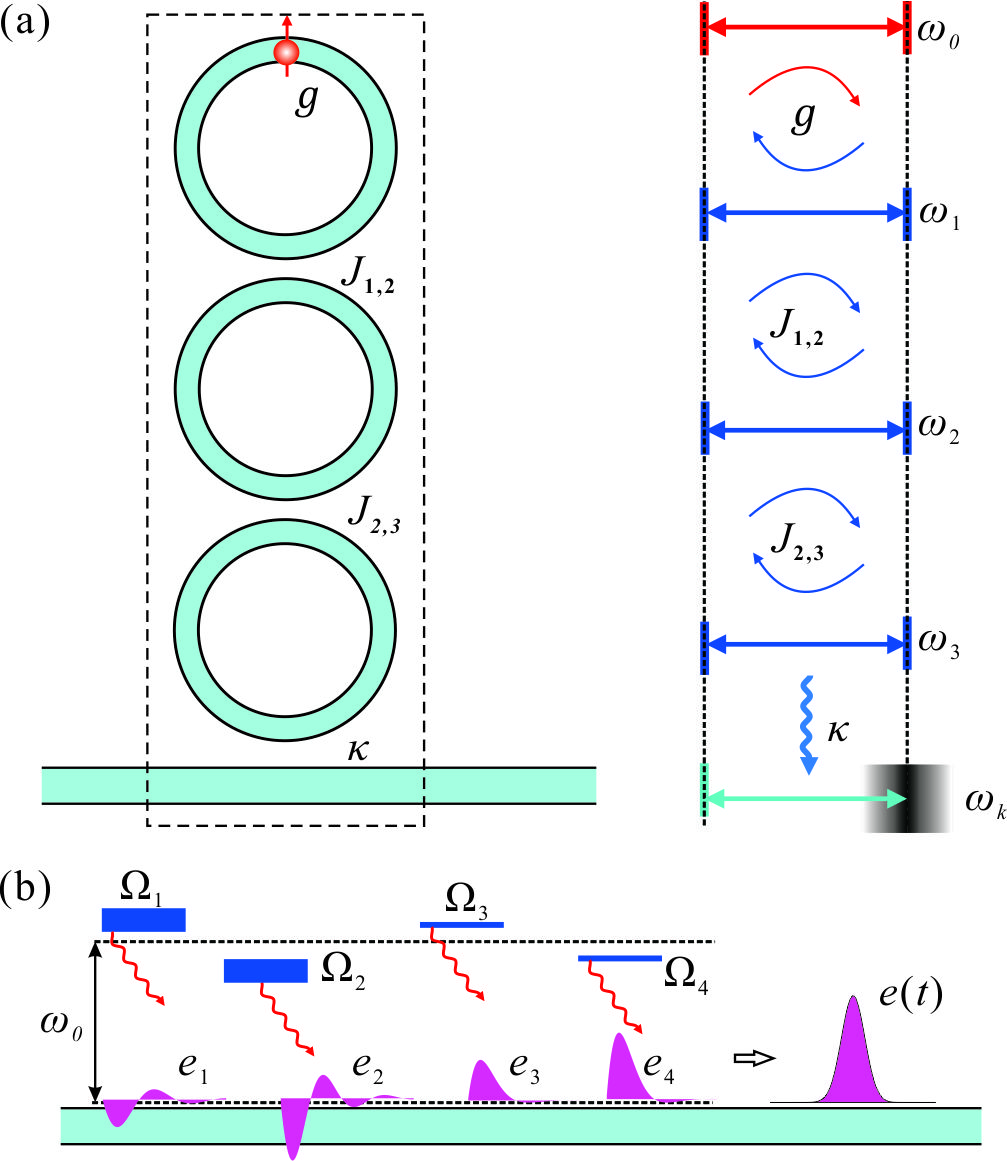}
      \caption{A quantum photonic node for on-chip state transfer. (a) Schematic diagram of the node (left) and coupling scheme (right). (b) Eigenstates of the node and the superposition picture of pulse shaping: $e(t)$ is the sum of the contributions from the eigenstates $e_{n}(t)$.}
      \label{fig1}
\end{figure}
\par Figure \ref{fig1}(a) schematically illustrates the proposed quantum node and its coupling schemes. We assume that the coupling of the TLS with the MRR modes is much greater than its original spontaneous decay rate $\Gamma_{0}$ and the radiation decay of the MRR modes $\Gamma_{c}$ is negligible compared to MRR-MRR and MRR-waveguide coupling rates, which guarantee the emission of the TLS couples predominantly into the waveguide. The assumptions are realistic with the consideration of current integrated photonics and solid-state quantum technologies \cite{RN107,RN108,RN120,RN121} and the possible implementation will be discussed later. The quantum photonic node can be analyzed in terms of the eigenstates of the whole coupled system and the time evolution of the single-photon probability amplitude in the last MRR coupling to the continuum can be considered as a result of superposition of all the eigenstates, as displayed in Fig. \ref{fig1}(b). Crucially, the pulse shape in the waveguide is a copy of that in the last MRR \cite{RN143}, which can be synthesized to time-reversal symmetric by optimizing various coupling rates, including TLS-MRR coupling constant $g$, MRR-MRR hopping rate $J_{n,n+1}$ and MRR-waveguide coupling rate $\kappa$, as indicated in Fig. \ref{fig1}(a).
\par The emission and excitation processes of the quantum node can be rigorously formulated by applying the waveguide and cavity quantum electrodynamics model \cite{RN15,RN99,RN98}. The system Hamiltonian reads as
\begin{equation}
\label{equation1}
	\hat{H}=\hat{H}_{0}+\hat{H}_{AM}+\hat{H}_{MM}+\hat{H}_{MW}
\end{equation}
consisting of uncoupled Hamiltonian  $\hat{H}_{0}/\hbar=\omega_{0}\hat{\sigma}_{+}\hat{\sigma}_{-}+\omega_{n}(\hat{a}_{n}^{\dagger}\hat{a}_{n}+\hat{b}_{n}^{\dagger}\hat{b}_{n})+\omega_{k}\hat{c}_{k}^{\dagger}\hat{c}_{k}$, the TLS-MRR coupling $\hat{H}_{AM}/\hbar=g(\hat{\sigma}_{+}\hat{a}_{1}+\hat{\sigma}_{+}\hat{b}_{1}+h.c.)$, MRR-MRR coupling $\hat{H}_{MM}/\hbar=J_{n,n+1}(\hat{a}_{n}^{\dagger}\hat{b}_{n+1}+\hat{b}_{n}^{\dagger}\hat{a}_{n+1}+h.c.)$ and MRR-waveguide coupling $\hat{H}_{MW}/\hbar=V_{k-}\hat{a}_{N}^{\dagger}\hat{c}_{k-}+V_{k+}\hat{b}_{N}^{\dagger}\hat{c}_{k+}+h.c.$. We have applied Einstein summation for appropriate terms over repeated indices, and the abbreviation $h.c$. for Hermitian conjugate. Here $\omega_{0}$ is the TLS transition frequency and $\hat{\sigma}_{\pm}$  the raising/lowering operator. The MRRs are described by the creation (annihilation) operators of clockwise and counterclockwise whispering gallery modes $\hat{a}_{n}^{\dagger} (\hat{a}_{n})$
 and $\hat{b}_{n}^{\dagger} (\hat{b}_{n})$, the resonant frequencies $\omega_{n}$ and the coupling rates of neighboring MRRs $J_{n,n+1}$, respectively. The TLS interacts only with the first MRR. The $N$-th MRR couples to the waveguide continuum denoted by the wave vector $k$ and its creation (annihilation) operator $\hat{c}_{k}^{\dagger}(\hat{c}_{k})$. For quantum state transfer, we consider only one excitation and express the state of the system in the interaction picture rotating at $\omega_{0}$ as 
\begin{equation}
\label{equation2}
	|\psi\rangle=[c_{0}(t)\hat{\sigma}^{+}+c_{n,a}(t)\hat{a}_{n}^{\dagger}+c_{n,b}(t)\hat{b}_{n}^{\dagger}+c_{k}(t)\hat{c}_{k}^{\dagger}]|\varnothing \rangle
\end{equation}
where $|\varnothing \rangle$ denotes the system state with no excitation. The time evolution of the probability amplitudes $c_{0},c_{n,a},c_{n,b}$ and $c_{k}$, can be determined by substituting Eq. \eqref{equation2} into the Schr\"{o}dinger equation $i\hbar\partial_{t}|\psi\rangle=\hat{\mathscr{V}}|\psi\rangle$, where $\hat{\mathscr{V}}$ is the interaction Hamiltonian deduced from $\hat{H}$. Considering the symmetry between the clockwise and counterclockwise modes, we drop the index $a(b)$ hereafter and define the probability amplitude $c_{n}=\sqrt{2}c_{n,a}=\sqrt{2}c_{n,b}$ so that $|c_n |^2$ is the probability for the photon residing in the $n$-th MRR. By applying Weisskopf-Wigner approximation to eliminate the continuum \cite{RN83}, we obtain a dynamic equation for the probability amplitude vector $\bm{c}=[c_{0},c_{1},c_{2}...,c_{N}]^{{\rm T}}$ in a matrix form as follows \cite{RN143}
\begin{equation}
\label{equation3}
	\partial_{t}\bm{c}=-i \bm{{\rm H}}\bm{c}+\bm{d}
\end{equation}
$\bm{{\rm H}}$ has a tridiagonal form $\bm{{\rm H}}=\text{tridiag}[\bm{u},\bm{v},\bm{u}]$ with $\bm{u}=(\sqrt{2}g,J_{12},J_{23},...,J_{N-1,N}),\bm{v}=(0,\delta_{1},\delta_{2},\delta_{3},...,\delta_{N})$,$\delta_{n}=\omega_{n}-\omega_{0} \quad (n=1,...,N-1)$, and $\delta_{N}=\omega_{N}-\omega_{0}-i\kappa/2$. The inhomogeneous term $\bm{d}$ serves as the drive for the system. The single-photon probability amplitude $e(t)$ in the waveguide is directly related to $c_{N}(t)$ as $e(t)=-i\sqrt{\kappa} c_{N}(t)$ \cite{RN143}.
\par We first discuss the process of transferring one excitation in the TLS into the waveguide. In this case $\bm{d}=0$, with the initial conditions of $c_{0}(0)=1$ and $c_{n}(0)=0$, Eq. \eqref{equation3} can be solved analytically. One has 
\begin{equation}
\label{equation4}
	e(t)=\sum_{n=1}^{N+1}\alpha_{n}e^{-i\Omega_{n}t}
\end{equation}
where $\Omega_{n}$ is the $n$-th complex eigenvalue of $\bm{{\rm H}}$ and $\alpha_{n}=-i\sqrt{2\kappa }g\prod_{m=1}^{N-1} J_{m,m+1}/\prod_{m,m\neq n}^{N+1}(\Omega_{n}-\Omega_{m})$ is the complex amplitude. One clearly sees from Eq. \eqref{equation4} that the node has $N+1$ eigenstates and the pulse shape is a result of superposition of all the eigenstates, confirming the physical picture of interference. By defining  $e_{n}(t)=\alpha_{n}e^{-i\Omega_{n}t}$, one sees that the pulse shape can be tuned by controlling the complex amplitudes and eigenvalues as schematically shown in Fig. \ref{fig1}(b). While the real part of $\Omega_{n}$ indicates energy shift relative to $\omega_{0}$, the imaginary part is due to coupling to the waveguide and means exponential decay of the eigenstate. Next we harness the interference effect to synthesize a time-symmetric single-photon wave packet. To quantify the pulse symmetry, we introduce a symmetry factor $\beta=\text{max}_{t_{0}}(\int_{-\infty}^{\infty}|e(t)e(2t_{0}-t)|dt)^2$, which is unity for a perfectly symmetric pulse. For the sake of conciseness, we consider here only the resonant case $\omega_{n}=\omega_{0}$. From Eq. \eqref{equation3} and the expression of $\bm{{\rm H}}$, one sees that only the ratios, i.e. $(J_{n,n+1},\kappa)/g$, affect the pulse shape. Although superposition from more eigenstates provides more degrees of freedom and promises better performance, we focus on the practical configuration of $N=3$. The results for $N=$ 1, 2 and $4$ are included in the Supplemental Material \cite{RN143}.
\begin{figure}
\centering
\includegraphics[width=8.6cm]{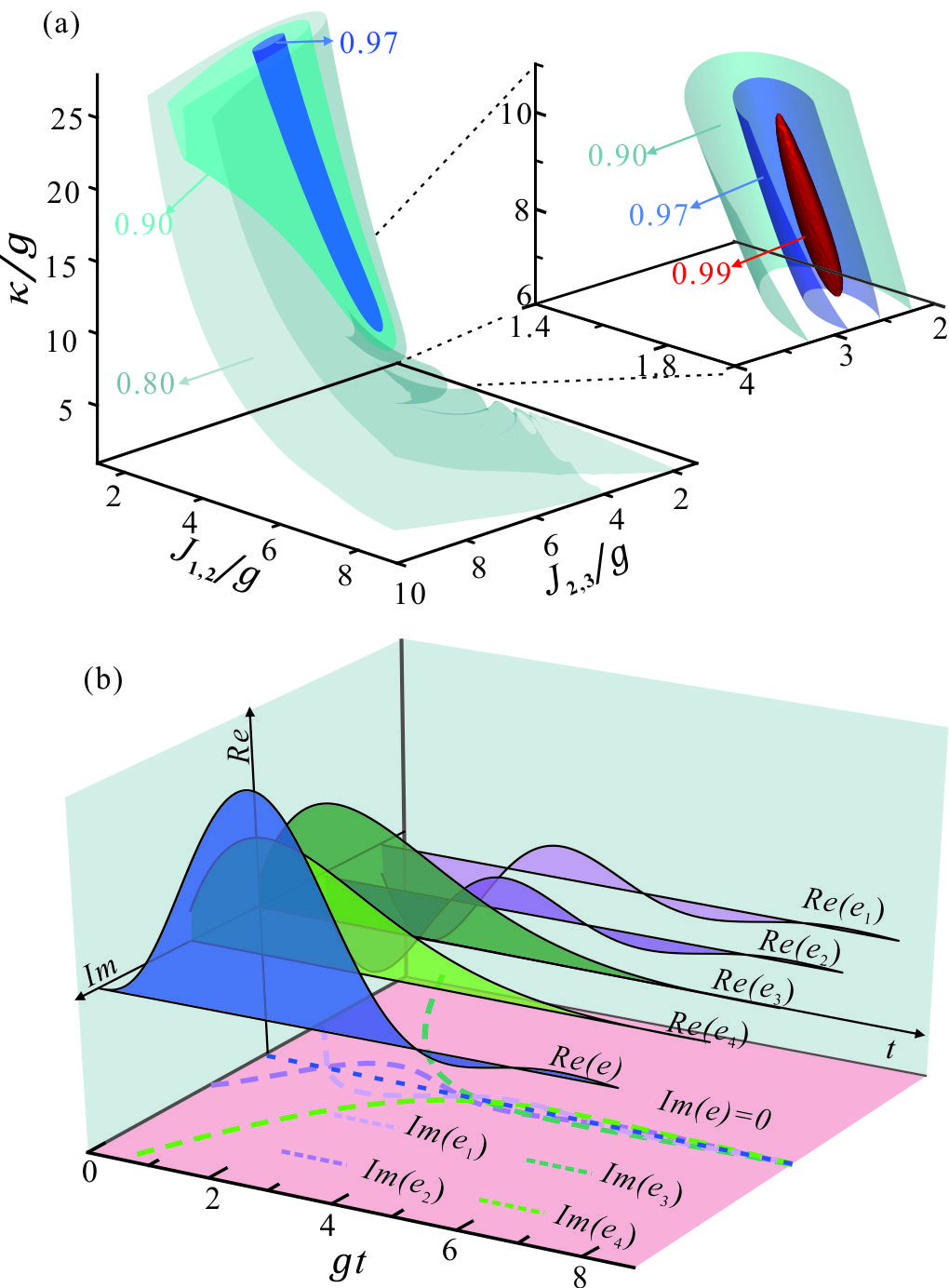}
\caption{(a) Time symmetric factor $\beta$ of the single-photon pulse emitted from the quantum node as a function of $(J_{12},J_{23},\kappa)/g$. (b) Pulse synthesis: the optimal pulse $e(t)$ with $(J_{12},J_{23},\kappa)/g=(1.88, 2.94, 7.92)$ expressed as a coherent superposition of emissions from four eigenstate channels.}
\label{fig2}
\end{figure}
\par Figure \ref{fig2}(a) displays $\beta$ in color-coded contours as a function of three normalized coupling rates $(J_{12},J_{23},\kappa)/g$. One observes that $\beta$ beyond 0.9 can be achieved in a pretty large parameter space. Part of the space is zoomed in and visualized with the contours of $\beta=0.9$, $0.97$, and $0.99$, highlighting that $\beta>$ 0.99 is achievable. In particular, the maximum value of $\beta=0.993$ is obtained for $(J_{12},J_{23},\kappa)/g=(1.88,2.94,7.92)$. We computed the eigenstates of the quantum node with the above optimal parameters and found that the system splits into two pairs of states with eigenvalues of $\Omega_{1,2}=(\pm 2.84-0.88i)g$ and $\Omega_{3,4}=(\pm 1.02-0.95i)g$, respectively. To intuitively understand the result about the symmetry factor, we plot in Fig. \ref{fig2}(b) the time evolutions of the complex probability amplitude $e(t)$ and the components $e_{n}(t)$ from the four independent emitting states.The imaginary and real parts are separately shown. In particular, the imaginary parts plotted in dashed traces are projected onto the bottom plane while the real parts are presented as traces with shaded areas in the planes laterally shifted with regard to the real axis plane. One observes that the imaginary part of each channel destructively cancels to zero and the real part adds up to the final more symmetric $e(t)$, although each individual emitting state shows Rabi-like oscillations with an exponential decay. The four different emitting states with distinct phases, amplitudes and Rabi frequencies synthesize a pulse with a near-unity symmetry factor.
\begin{figure}
\centering
\includegraphics[width=8.6cm]{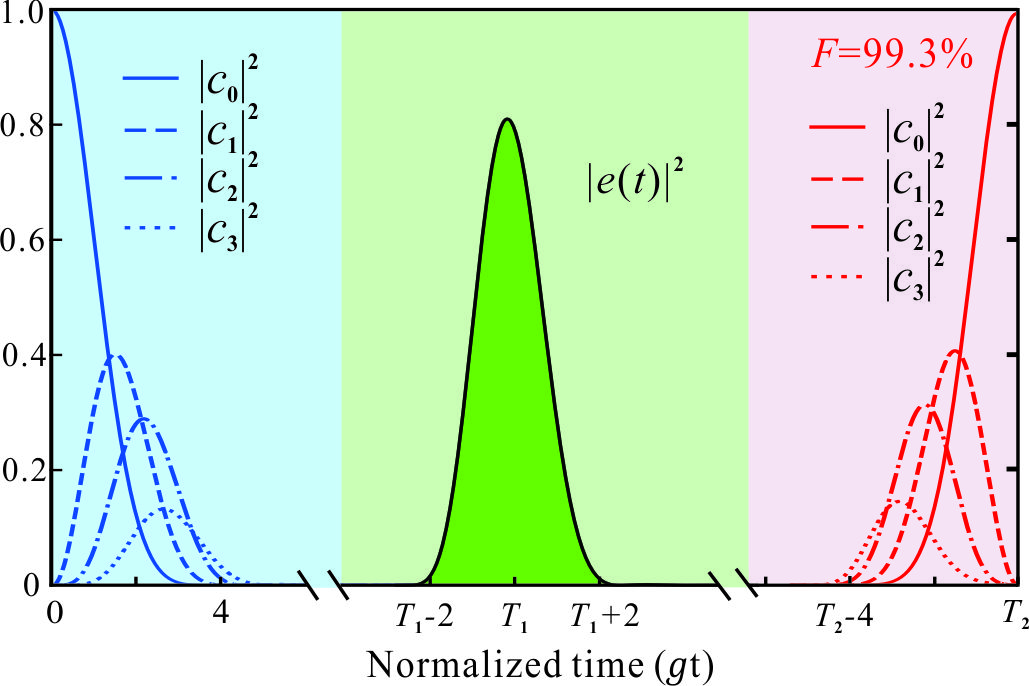}
\caption{Complete quantum state transfer process from one quantum node to another. Time evolution of excitation probabilities of various channels in the sending (light blue), transport (light green) and receiving sections (pinkish shaded).}
\label{fig3}
\end{figure}
\par We have thoroughly studied the process of sending one excitation of a node into the waveguide a time symmetric single-photon wave packet. In the following, we examine the receiving process. With an arbitrary incoming single-photon probability amplitude $f(t)$ in the waveguide, the dynamics of the receiving process $c_{0}(t)$ and $c_{n}(t)$ can be obtained by solving Eq. \eqref{equation3} with a drive term $\bm{d}=[0,0,...,-i\sqrt{\kappa}f(t)]^{{\rm T}}$. We assume the receiving process ends when the population of the TLS $|c_{0}(t)|^2$ reaches the maximum. Our modeling allows access to the whole quantum state transfer dynamic process, including the emission process with one excitation in the TLS, the propagation of single-photon packet through the waveguide and the excitation of the receiving node with the single-photon wave packet $f(t)=e(t)$. To evaluate the performance of state transfer, we define the maximum population of the TLS in the receiving node as $F$, which is also the overall success rate. Figure \ref{fig3} presents the dynamics of the whole quantum state transfer process between two quantum nodes with $N=3$ and with the optimal coupling parameters obtained in Fig. \ref{fig2}(a). The time evolutions of the emission, transport and absorption processes are demonstrated with various line traces in the light blue, green and pinkish shaded regions, respectively. As expected, the single-photon wave packet in the waveguide is highly symmetric in time and thus can excite the identical quantum node with maximum probability according to the time reversal symmetry \cite{RN17,RN37}. One also clearly observes the time reversal symmetry of the sending and the receiving nodes. With the optimal coupling rates for the node with $N=3$, an overall success rate of $F=0.993$ (equal to the symmetry factor) can be achieved.
 \begin{figure}
  \centering
      \includegraphics[width=8.6cm]{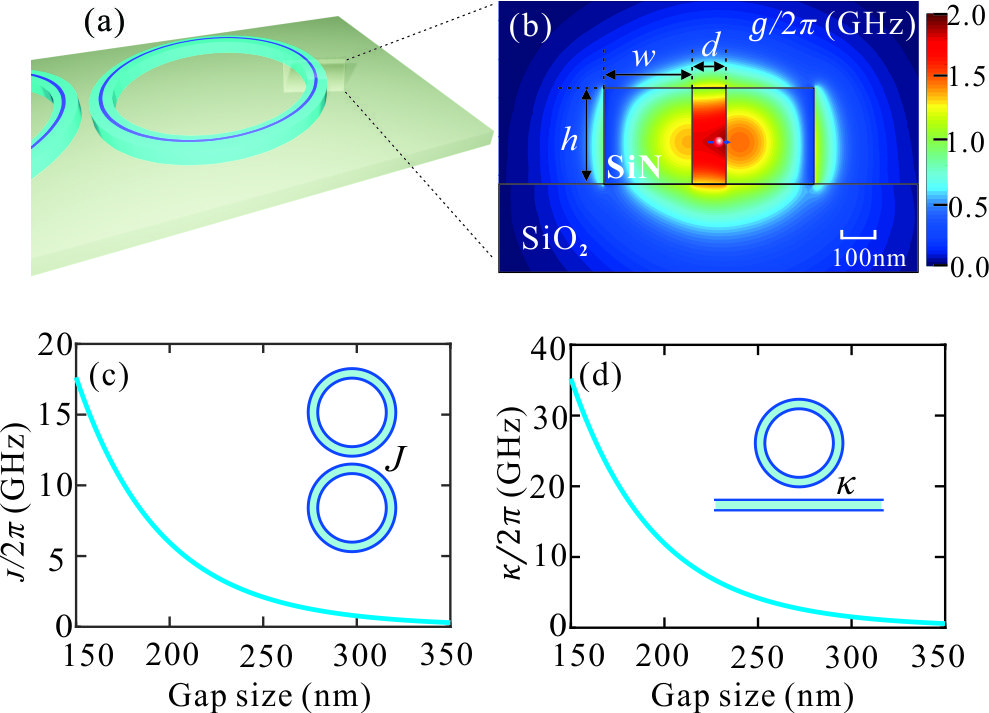}
  \caption{Possible experimental realization of the proposed quantum photonic node. (a) A sketch of the node based on SiN slot waveguide on silica and single molecules as quantum emitters. (b) TLS-MRR coupling constant $g$ overlaid on the cross section of the slot waveguide. (c) MRR-MRR coupling rate $J$ and (d) MRR-waveguide coupling rate $\kappa$ as functions of the gap size.}
   \label{fig4}
\end{figure}
\par Next we discuss and provide guidelines for the experimental realization of such quantum photonic nodes with integrated photonic platforms and solid-state quantum optics techniques. As a concrete example, here we explore the possibility of silicon nitride on silica platform for building photonic circuits and single organic molecules such as dibenzoterrylene (DBT) molecules in anthracene matrix \cite{RN130,RN129} as solid-state emitters. Specifically, as schematically displayed in Fig. \ref{fig4}(a) and \ref{fig4}(b), one may fabricate MRRs based on silicon nitride (SiN) slot waveguide with a radius of about 10 ${\rm \mu m}$. The rectangular slot waveguide cross section is specified by $(w,d,h)=(250,100,250)$ nm and the slot can be filled by anthracene matrices with single molecules embedded \cite{RN129,RN133}. The platform of SiN on silica has the advantages of being transparent in visible and near-infrared range and having achievable waveguide loss rate as low as 0.1 dB/m \cite{RN132}. Such kind of MRR structures could provide intrinsic quality factors in the order of $10^7 (\Gamma_{c}\sim\text{30 MHz})$. The TLS-MRR coupling constant $g$ can be calculated according to $g=0.5\sqrt{3\lambda^2c\Gamma_{0}/(2\pi n^3 V_{{\rm eff}})}$, where $c,n,V_{{\rm eff}}$ are the speed of light, refractive index and the effective mode volume of the mode \cite{RN14}, respectively. Figure \ref{fig4}(b) shows a color-coded contour map of $g$ on the cross section of the waveguide for DBT molecules with natural linewidths of about 30 MHz. Coupling constants of a few GHz is achievable. For the $N=3$ node with the optimal parameters of  $(J_{12},J_{23},\kappa)/g=(1.88,2.94,7.92)$, the emission of the molecules will be delivered to the waveguide with efficiencies up to about 99\%. Based on the above basic configuration, we examine the achievable parameter ranges of MRR-MRR coupling rate $J$ and MRR-waveguide coupling rate $\kappa$, which can be calculated semi-analytically \cite{RN143,RN32}. Figure \ref{fig4}(c) and \ref{fig4}(d) depict the coupling rate $J$ between two MRRs and the decay rate $\kappa$ of MRR to the waveguide as a function of MRR-MRR and MRR-waveguide gap sizes, respectively. The TLS-MRR coupling constant $g$ is only slightly modified due to the small changes of $V_{{\rm eff}}$ in the presence of coupling. One clearly observes that the required $J$ and $\kappa$ for realizing the optimal quantum node are feasible by tuning the gap size and their values are orders of magnitude greater than the radiation loss of the MRRs. The guidelines and the designs given here are to demonstrate the experimental feasibility for nearly perfect quantum state transfer on chip. Alternative cavity structures, emitter systems and design strategies should certainly be explored for other experimental realizations.  In the Supplemental Material \cite{RN143}, we discuss the effects of various non-deal conditions on the transfer success rate, such as the influences of considering the spontaneous emission, cavity loss, TLS-MRR resonance detuning and cross coupling between clockwise and counter clockwise whispering gallery modes.
\par We have proposed and carefully studied a type of full waveguide-structure based quantum photonic node capable of performing deterministic quantum state transfer between distant nodes via propagating photons with near-unity overall success rate. Our scheme doesn't require any type of dynamic modulation and can be implemented in CMOS-compatible integrated photonic platforms \cite{RN108,RN120,RN121}, promising its experimental realization in near future. We essentially devised an approach that can completely transfer a dipolar excitation to another dipolar system through propagating wave packet synthesis. The format of the dipolar excitation is not limited only to optical two-level systems but could be extended to other systems, for instance superconducting qubit systems \cite{RN103,RN104} and optomechanical systems \cite{RN146}, which are highly promising alternative platform for quantum information processing. It should also be applicable to hybrid quantum systems such as nitrogen vacancy centers and superconducting qubits \cite{RN118} or hybrid acoustic and superconducting qubits \cite{RN117}. We believe our work paves the way for on-chip state transfer in various quantum and classical network systems.

\begin{acknowledgments}
We acknowledge financial support from the National Natural Science Foundation of China (Grant Number 11874166, 11604109, 11474114), the Thousand-Young-Talent Program of China and Huazhong University of Science and Technology. X.-W.C would like to thank Vahid Sandoghdar for inspiring discussions, continuous support and encouragement. 
\end{acknowledgments}

\textbf{CONFLICT OF INTEREST: } X.-W.C, P.Z and Z.T have filed a patent application based on the present work on 06.08.19 (Patent application number 201910722382.1). 

\bibliographystyle{apsrev4-1}

\end{document}